\documentclass[pra,showpacs,bibnotes]{revtex4}
\usepackage{graphicx}
\begin{document}
\draft
\def\ds{\displaystyle}
\title{Dynamical Casimir Effect for a Swinging Cavity }
\author{Cem Yuce$^{*}$, Zalihe Ozcakmakli}
\address{Department of Physics, Anadolu University,
Eskisehir, Turkey.\\ cyuce@anadolu.edu.tr}
\email{cyuce@anadolu.edu.tr}
\begin{abstract}
The resonant scalar particle generation for a swinging cavity
resonator in the Casimir vacuum is examined. It is shown that the
number of particles grows exponentially when the cavity rotates at
some specific external frequency.
\end{abstract}
\maketitle

\section{Introduction}

In 1948, Casimir predicted an attractive force between two closely
placed perfectly conducting plates due to the vacuum fluctuations
\cite{z1}. The recent measurements confirm the basic concepts of
quantum field theory in the presence of static external
constraints \cite{deneyCas}. The best way to learn more about the
quantum vacuum is to distort the vacuum. One way to distort the
vacuum is to vibrate one of the wall with one of the resonant
field frequencies. Under this resonance condition, the vacuum
field and the moving mirrors become strongly coupled to each
other. The time dependent boundary conditions due to the moving
mirrors chance the vacuum field in the cavity. As a result, it was
theoretically predicted that photons are generated in the empty
cavity because of the instability of the vacuum state of the
electromagnetic field in the presence of time-dependent boundary
conditions \cite{z2,z3,z4,z31,ekle1,ekle2,ekle3,z9,z11}. A number
of virtual photons from the vacuum are converted into real
photons. This phenomenon is known as dynamical Casimir effect or
motion-induced radiation. There are a few proposed experimental
for the detection of photons \cite{qw1,qw2,deney1}.\\
A one dimensional cavity with two perfectly parallel reflecting
walls, one of which is motionless and the other oscillating with a
mechanical frequency equal to a multiple of the fundamental
optical resonance frequency of the static cavity, has been used as
a simple model to study the dynamical Casimir effect
\cite{z3,z4,z9,z11}.
\begin{equation}\label{L1L2gdh}
L_{1}(t) = 0~,~~ L_{2}(t) ~= L\left(1+\epsilon ~ \sin \Omega
t\right)~ ,
\end{equation}
where the constant $\Omega $ is the external frequency,
$\ds{\epsilon}$ is a small parameter, the constant $\ds{L}$ is the
initial length of the cavity and $\ds{L_{1}(t)}$ and
$\ds{L_{2}(t)}$ are the positions of the right and the left walls
at time $\ds{t}$, respectively. The cavity is motionless initially
and that at some instant one mirror starts to oscillate resonantly
with a tiny amplitude. The case of cavities with two moving
mirrors was studied by few authors
\cite{z16,z17,twomoving1,twomoving2,twomoving3,twomoving4,twomoving5,z5,yuceozcak}.
Compared to the situation with a single oscillating mirror, it was
found that radiation is resonantly enhanced when the cavity
oscillates symmetrically with respect to the cavity center. It was
also shown that photons are generated when the cavity oscillates
as a whole, that is, the length of the cavity is constant \cite{yuceozcak}. \\
So far the attention has been mainly focused on the oscillation of
the mirrors of the cavity. Among the previous works on the quantum
vacuum, there is a lack of information about the resonance photon
generation for a rotating cavity in vacuum. In this paper, we will
examine the resonance scalar particle generation when the cavity
rotates around itself. To our knowledge, particle generation due
to the rotation of the cavity was not considered before in the
literature.

\section{Scalar Field}

Consider a cubic cavity formed by perfect conductors whose sides
have a length $L$. The scalar field vanishes on the surfaces. So,
the Dirichlet boundary conditions read
\begin{equation}\label{bcnormal}
\Phi(x=\{0,L\},y,z,t)=0~,~~\Phi(x,y=\{0,L\},z,t)=0~,~~\Phi(x,y,z=\{0,L\},t)=0~,
\end{equation}
where the field operator in the Heisenberg representation obeys
the wave equation, $c=1$
\begin{equation}\label{0z123hgk}
\nabla^{2}\Phi=\frac{\partial^{2}\Phi}{\partial t^{2}}~.
\end{equation}
The field operator inside the cavity can be expanded as
\begin{equation}\label{expansion}
\Phi(\mathbf{x},t)=\sum_{n}~\left(~b_{\mathbf{n}}\Psi_{\mathbf{n}}(\mathbf{x},
t)+b_{\mathbf{n}}^{\dag}\Psi_{\mathbf{n}}^{\star}(\mathbf{x},
t)~\right)~,
\end{equation}
where $\ds{b_{\mathbf{n}}^{\dag}}$and $\ds{b_{\mathbf{n}}}$ are
the creation and the annihilation operators, respectively and
$\ds{\Psi_{\mathbf{n}}(\mathbf{x}, t)}$ is the corresponding mode
function. For $t<0$, the mode function satisfying the boundary
conditions is expanded as
\begin{equation}\label{denklem0n0}
\Psi_{\mathbf{n}}=N_{\mathbf{n}}~e^{-i\omega_{\mathbf{n}}t}
~\sin(\frac{n_x\pi x}{L})\sin(\frac{n_y\pi y}{L})\sin(\frac{n_z\pi
z }{L})
\end{equation}
where $\omega_{\mathbf{n}}=\pi/L \sqrt{n_x^2+n_y^2+n_z^2}$ and the
constant
$\ds{N_{\mathbf{n}}=({2/L})^{\frac{3}{2}}(~1/\sqrt{2\omega_{\mathbf{n}}}~)
}$. \\
At time $t=0$, the cavity starts to rotate. In the literature, the
most common and useful way to describe the rotation of a rigid
body is to use the Euler angles. For the simplicity, assume that
the $z$-axis lies along the axis of rotation. The problem looks
simple, since the wave equation is the same as whether the
boundaries are rotating or not. However, the boundary conditions
become time-dependent. Although the solution of the wave equation
is easy to be found and well known in static case
(\ref{denklem0n0}), finding the exact solution of the same problem
endowed with the time-dependent boundary
conditions is very difficult.\\
In the following, we denote the fixed system in unprimed
coordinates and the rotated one in primed coordinates. The
orthogonal transformation matrix about the $z$-axis through the
angle $\theta$ transforms the fixed unprimed axes to the primed
axes leaving the distance between two points unchanged.
\begin{eqnarray}\label{z123hgk}
\left(%
\begin{array}{c}
  x^{\prime}\\
  y^{\prime}\\
  z^{\prime}\\
\end{array}%
\right)=\left(%
\begin{array}{ccc}
  ~~\cos~\theta (t)  & \sin~\theta (t)  & 0 \\
  -\sin~\theta (t)  & \cos~\theta (t)  & 0 \\
  0 & 0 & 1\\
\end{array}%
\right)\left(%
\begin{array}{c}
  x \\
  y \\
  z \\
\end{array}%
\right)
\end{eqnarray}
where $\dot{\theta}$ is the angular velocity and dot denotes time
derivation with respect to $t$.\\
The boundary conditions in the rotating system is given by
\begin{equation}\label{bcnormal2}
\Phi(x^{\prime}=\{0,L\},y^{\prime},z^{\prime},t)=0,~~\Phi(x^{\prime},y^{\prime}=\{0,L\},z^{\prime},t)=0,
~~\Phi(x^{\prime},y^{\prime},z^{\prime}=\{0,L\},t)=0~.
\end{equation}
One can also derive the time dependent boundary conditions by
using
(\ref{z123hgk}).\\
We look for the rotations of the cavity that could enhance the
number of generated photons by means of resonance effects for some
specific external frequencies. To enhance the number of photons,
let us now assume the resonance condition,
\begin{equation}\label{theta}
\theta (t) =\epsilon ~\sin{(\Omega ~t)}~,
\end{equation}
where the constant $\Omega$ is the external frequency and the
constant $\epsilon$ is a small parameter ($\ds{~L \dot{\theta}}$
is very small compared to the
speed of the light, where $L$ is the length).\\
There are two ways to deal with this problem. One can either solve
the wave equation with the time dependent boundary conditions or
transform the time dependent boundary conditions to the fixed
boundary conditions. In the latter case, the form of the wave
equation is changed. We will follow the second case. Let us now
find the new form of the wave equation. The derivative operators
transform according to
$\ds{\nabla^{2}\rightarrow{\nabla^{\prime}}^{2}}$,
$\ds{\frac{\partial}{\partial
t}\rightarrow\frac{\partial}{\partial t}-i\dot{\theta}J_z}$ where
$\ds{iJ_z=x^\prime\frac{\partial}{\partial
y^\prime}-y^\prime\frac{\partial}{\partial x^\prime}}$.
Substituting these into the wave equation (\ref{0z123hgk}) and
neglecting the terms with $\ds{\dot{\theta}^2}$ gives the
transformed wave equation in the primed coordinate system.
\begin{equation}\label{z123}
{\nabla^{\prime}}^2\Phi=\frac{\partial^2\Phi}{\partial
t^2}-iJ_z(2~\dot{\theta}~\frac{\partial}{\partial
t}+\ddot{\theta})~\Phi~.
\end{equation}
The last term in the right hand side is due to the rotation of the
cavity. Now, we do not have to deal with the time dependent
boundary conditions. However, we are left with a new equation.\\
We will try to solve this equation. To find the explicit form of
the field operator, we will follow the approach given in
\cite{z20,z21,z22}. For an arbitrary moment of time $t>0$, the
mode function (\ref{expansion}) satisfying the boundary conditions
(\ref{bcnormal2}) is expanded as
\begin{equation}\label{edywau}
\Psi_{\mathbf{k}}=\sum_{\mathbf{n}}c_{\mathbf{n}}^{\mathbf{k}}(t)~\left(\frac{2}{L}\right)^{\frac{3}{2}}\sin(\frac{n_x\pi
x^\prime}{L})\sin(\frac{n_y\pi y^\prime}{L})\sin(\frac{n_z\pi
z^\prime}{L})~,
\end{equation}
where
$\ds{c_{\mathbf{n}}^{\mathbf{k}}(t)=c_{n_xn_yn_z}^{k_xk_yk_z}(t)}$
is a time-dependent function to be determined later. The initial
conditions are given by
$\ds{c_{\mathbf{n}}^{\mathbf{k}}(0)=1/\sqrt{2\omega_{\mathbf{k}}}~\delta_{\mathbf{n},\mathbf{k}}}$
and
$\ds{\dot{c}_{\mathbf{n}}^{\mathbf{k}}(0)=-i\sqrt{\omega_{\mathbf{k}}/2}~\delta_{\mathbf{n},\mathbf{k}}}$.\\
Let us substitute (\ref{edywau}) into the equation (\ref{z123}) in
order to find $\ds{c_{\mathbf{n}}^{\mathbf{k}}(t)}$. If we
multiply the resulting equation with $~\ds{\sin(m_x\pi
x^\prime/L)\sin(m_y\pi y^\prime/L)\sin(m_z\pi z^\prime/L)}$ and
use the orthogonality relations, we get an infinite set of coupled
differential equations for $\ds{c^{\mathbf{k}}_{\mathbf{n}}}$
after some algebra.
\begin{equation}\label{denklem005}
\ddot{c}^{~\mathbf{k}}_{\mathbf{m}}+\omega^2_{\mathbf{m}}
~c^{\mathbf{k}}_{\mathbf{m}}=\epsilon ~\delta_{n_zm_z}\left(8
\sum_{n_x}\sum_{n_y} {g_{n_ym_y}^{n_xm_x}}
~\Theta^{\mathbf{k}}_{\mathbf{n}} +
\sum_{n_x}g_{n_xm_x}\Theta^{\mathbf{k}}_{\mathbf{n}}~\delta_{n_ym_y}-\sum_{n_y}g_{n_ym_y}\Theta^{\mathbf{k}}_{\mathbf{n}}~\delta_{n_xm_x}\right)~,
\end{equation}
where $\ds{g_{nm}=\frac{(1-(-1)^{m+n})mn}{m^2-n^2}}$, $\ds{~~
g_{n_ym_y}^{ n_xm_x}=\frac{1}{\pi^2}(\frac{1}{m_x^2-n_x^2}-
\frac{1}{m_y^2-n_y^2 } )~g_{n_xm_x}g_{n_ym_y}}$ and they are equal
to zero when $n_x= m_x$ or $n_y= m_y$.
$\Theta^{\mathbf{k}}_{\mathbf{n}}$ is defined as
\begin{equation}\label{FE}
\epsilon~
\Theta^{\mathbf{k}}_{\mathbf{n}}=\left(2~\dot{\theta}~\dot{c}^{~\mathbf{k}}_{\mathbf{n}}
+\ddot{\theta}~c^{\mathbf{k}}_{\mathbf{n}}\right)~.
\end{equation}
We will try to solve the second order differential equation
(\ref{denklem005}). Conventional weak-coupling perturbation theory
suffers from problems that arise from resonant terms in the
perturbation series. The effects of the resonance could be
insignificant on short time scales but become important on long
time scales. Perturbation methods generally break down after small
time whenever there is a resonance that lead to what are called
secular terms. In the equation (\ref{denklem005}), this happens
for those particular values of external frequency $\Omega$ such
that there is a resonant coupling with the eigenfrequencies of the
static cavity. A number of approaches have been developed to solve
such equations. For example, averaging over fast oscillations
\cite{z24,z26}, multiple scale analysis \cite{z27,z30}, the
rotating wave approximation \cite{z28}, numerical techniques
\cite{z29}. We prefer to use the multiple scale analysis method
(MSA), a powerful and sophisticated
perturbative method valid for longer times. \\
In MSA, the trick is to introduce a new variable
$\ds{\tau=\epsilon t}$. This variable is called the slow time
because it does not become significant in the small time. The
functional dependence of $\ds{c^{\mathbf{k}}_{\mathbf{m}}}$ on $t$
and $\epsilon$ is not disjoint because it depends on the
combination of $\epsilon t$ as well as on the individual $t$ and
$\epsilon$. The time variables $t$ and $\ds{\tau}$ are treated
independently in MSA. Thus, in place of
$\ds{c^{\mathbf{k}}_{\mathbf{m}}(t)}$, we write
$\ds{c^{\mathbf{k}}_{\mathbf{m}}(t,\epsilon t)}$. Let us expand
$\ds{c^{\mathbf{k}}_{\mathbf{m}}}$ in the form of a power series
in $\epsilon$
\begin{equation}
c^{\mathbf{k}}_{\mathbf{m}}(t)=c^{\mathbf{k}(0)}_{\mathbf{m}}(t,\tau)~+~\epsilon~c^{\mathbf{k}(1)}_{\mathbf{m}}(t,\tau)~+~O(\epsilon^2)\\
\end{equation}
Up to the first order of $\epsilon$, the derivatives with respect
to the time scale $\ds{t}$ are given by
\begin{eqnarray}\label{msa2}
\dot{c}^{~\mathbf{k}}_{\mathbf{m}}&=& \partial_t~
c^{\mathbf{k}(0)}_{\mathbf{m}}+\epsilon~\left(\partial_{\tau}c^{\mathbf{k}(0)}_{\mathbf{m}}+\partial_{t}c^{\mathbf{k}(1)}_{\mathbf{m}}\right)\nonumber\\
\ddot{c}^{~\mathbf{k}}_{\mathbf{m}} &=&
\partial_{t}^2~c^{\mathbf{k}(0)}_{\mathbf{m}}+\epsilon~\left(2\partial_{\tau
}\partial_{
t}~c^{\mathbf{k}(0)}_{\mathbf{m}}+\partial_{t}^2~c^{\mathbf{k}(1)}_{\mathbf{m}}\right)~.
\end{eqnarray}
Let us substitute these into the equation (\ref{denklem005}).
Then, we see that our original ordinary differential equation is
replaced by a partial differential equation. It may appear that
the problem has been complicated. But, as will be seen below,
there are many advantages of this method. For the zeroth order of
$\epsilon$, we obtain a well known equation:
\begin{equation}
\ddot{c}_{\mathbf{m}}^{~\mathbf{k}(0)}+\omega^2_{\mathbf{m}}~
c_{\mathbf{m}}^{\mathbf{k} (0)}=0~.
\end{equation}
The solution is given by
\begin{equation}\label{p2qwtefirst}
c^{\mathbf{k}(0)}_{\mathbf{m}}=B^{\mathbf{k}}_{\mathbf{m}}(\tau)~e^{i\omega_{\mathbf{m}}~t}+C^{\mathbf{k}}_{\mathbf{m}}(\tau)~e^{-i\omega_{\mathbf{m}}~t}~,
\end{equation}
where
$B^{\mathbf{k}}_{\mathbf{m}}(\tau)=B_{m_xm_ym_z}^{k_xk_yk_z}(\tau)$,
$C^{\mathbf{k}}_{\mathbf{m}}(\tau)=C_{m_xm_ym_z}^{k_xk_yk_z}(\tau)$
are small time varying functions. The initial conditions are given
by
\begin{equation}\label{initialcond}
C^{\mathbf{k}}_{\mathbf{m}}(\tau=0)=\frac{1}{\sqrt{2\omega_{{\mathbf{k}}}}}~\delta_{\mathbf{k},\mathbf{m}}~;~~~~
B^{\mathbf{k}}_{\mathbf{m}}(\tau=0)=0~.
\end{equation}
To first order in $\epsilon$, we get
\begin{equation}\label{p2qwtefirst2}
\ddot{c}^{~\mathbf{k}(1)}_{\mathbf{m}}+\omega^2_{\mathbf{m}}
c^{\mathbf{k}(1)}_{\mathbf{m}}=\delta_{n_zm_z}\left(8
\sum_{n_x}\sum_{n_y}g_{n_ym_y}^{n_xm_x}~\Theta^{\mathbf{k}(0)}_{\mathbf{n}}
+\sum_{n_x}g_{n_xm_x}\Theta^{\mathbf{k}(0)}_{\mathbf{n}}
\delta_{n_ym_y}-\sum_{n_y}g_{n_ym_y}\Theta^{\mathbf{k}(0)}_{\mathbf{n}}
\delta_{n_xm_x}\right)-2\partial_t\partial_\tau~c_{\mathbf{m}}^{\mathbf{k}(0)}
\end{equation}
Let us substitute the definition (\ref{FE}) with (\ref{theta}) and
the zeroth order solution (\ref{p2qwtefirst}) into the right hand
side of the equation (\ref{p2qwtefirst2}) and then use the
following relations: $\ds{~2i\sin \Omega t=(e^{i\Omega
t}-e^{-i\Omega t})}$, $\ds{~~2\cos \Omega t=(e^{i\Omega
t}+e^{-i\Omega t})}$. It can be seen that the right hand side
contains terms that produce secular terms. For a uniform
expansion, these secular terms must vanish. In other words, any
term with $\ds {~e^{\pm i\omega_{\mathbf{m}}t}~}$ on the
right-hand side must vanish. If not, these terms would be in
resonance with the left-hand side term and secularities would
appear. After imposing the requirement that no term ~$\ds {e^{\mp
i\omega_{\mathbf{m}}t}~}$ appear, we get the equations for
$B_{\mathbf{m}}^{\mathbf{k}}(\tau)$,~~
$C_{\mathbf{m}}^{\mathbf{k}}(\tau)$. Although there are infinitely
many terms in the summations, only a few modes are coupled to each
other because of the nonequidistant character of the spectrum.
There are only a few positive integers $\ds{n_{x}, n_{y}}$ that
produce secular terms in (\ref{p2qwtefirst2}) for a given external frequency $\Omega$.\\
As an illustration, we will give a specific example. Assume that
the inter-mode coupling occurs between $\ds{(1,1,1)}$ and
$\ds{(1,2,1)}$, $\ds{(2,1,1)}$. Note that the external frequency
$\Omega$ should be determined in such a way that these modes are
coupled to each other. So, choose
$\ds{\Omega=(\sqrt{3}+\sqrt{6})\pi/L}$. As a result, a set of
equations for $B^{\mathbf{k}}_{\mathbf{m}}$ and
$C^{\mathbf{k}}_{\mathbf{m}}$ for these modes are given by
\begin{eqnarray}
  \partial_{\tau} B^{k_xk_yk_z}_{111}-G^{111}_{121}~C^{k_xk_yk_z}_{121}+G^{111}_{211}~C^{k_xk_yk_z}_{211}&=&0~, \nonumber\\
  \partial_{\tau} C^{k_xk_yk_z}_{111}-G^{111}_{121}~B^{k_xk_yk_z}_{121}+G^{111}_{211}~B^{k_xk_yk_z}_{211}&=&0~, \nonumber\\
  \partial_{\tau} B^{k_xk_yk_z}_{121}-G_{111}^{121}~C^{k_xk_yk_z}_{111}&=&0~, \nonumber\\
  \partial_{\tau} C^{k_xk_yk_z}_{121}-G_{111}^{121}~B^{k_xk_yk_z}_{111}&=&0~, \nonumber\\
  \partial_{\tau}B^{k_xk_yk_z}_{211}+G^{211}_{111}~C^{k_xk_yk_z}_{111}&=&0~,  \nonumber\\
  \partial_{\tau}C^{k_xk_yk_z}_{211}+G^{211}_{111}~B^{k_xk_yk_z}_{111}&=&0~,
\end{eqnarray}
where
$\ds{~G_{\mathbf{m}}^{\mathbf{k}}=~\frac{(2\omega_{\mathbf{m}}-\Omega)~\Omega}{4\omega_{\mathbf{k}}}~~g_{m_ik_i}~}$
($i=x$ if $\ds{m_x\neq k_x}$ or $i=y$ if $\ds{m_y\neq k_y}$).\\
The solutions can be obtained by solving the above six equations
with the initial conditions (\ref{initialcond}). The solutions for
$B^{\mathbf{k}}_{\mathbf{m}}(\tau)$ is given by
\begin{eqnarray}
\left(%
\begin{array}{c}
 B_{111}^{211} \\
 B_{111}^{121} \\
 B_{121}^{111} \\
 B_{211}^{111} \\
\end{array}%
\right)=
\frac{1}{\lambda}\left(%
\begin{array}{c}
  -G^{111}_{211}/\sqrt{2\omega_{211}}\\
  ~~G^{111}_{121}/\sqrt{2\omega_{121}}\\
  ~~G^{121}_{111}/\sqrt{2\omega_{111}}\\
  -G^{211}_{111}/\sqrt{2\omega_{111}}\\
\end{array}%
\right)~\sinh(\lambda\tau)~.
\end{eqnarray}
where
$\lambda^2=G^{111}_{121}G^{121}_{111}+G^{111}_{211}G^{211}_{111}$.\\
Assume that the cavity returns to its initial orientation at time
$t_{f}$ and stops rotating. The number of generated scalar
particles in the mode $(m_x,m_y,m_z)$ are calculated using the
following formula \cite{z30,z31}.
\begin{equation}
<N_{\mathbf{m}}>=\sum_{\mathbf{k}}2~\omega_{\mathbf{m}}~|
B_{\mathbf{m}}^{\mathbf{k}}|^{2}~.
\end{equation}
Hence, we find the average number of scalar particles generated in
the modes $(1,1,1)$, $(1,2,1)$ and $(2,1,1)$.
\begin{equation}\label{eyusty}
<N_{111}>=2<N_{121}>=2<N_{211}>=\sinh^{2}(\lambda~\tau_{f})~.
\end{equation}
They increase exponentially in time. The similar result has been
obtained for the dynamical Casimir effect problem \cite{z27,z30}.
There it was assumed that the mirrors of the cavity oscillate in
time.

\section{Discussion}

So far in the literature, the dynamical Casimir effect problem has
been studied only for the cavity whose mirror(s) is(are) in
oscillatory motion. Here, we have investigated the same effect for
the cavity which rotates around itself. It was theoretically
suggested that the particles are generated from the rectangular
cavity whose mirror is moving. Here we have shown that the
particles are generated from the rotating rectangular cavity as well.\\
If the cavity is rotated at a constant acceleration, no resonance
effect takes place. Resonance particle generation occurs if the
angle $\theta(t)$ changes sinusoidally (\ref{theta}). For the
translational dynamical Casimir effect problem, that occurs when
the mirror accelerates sinusoidally (\ref{L1L2gdh}).\\
The boundary conditions become time-dependent as the cavity
rotates. We have transformed the time-dependent boundary
conditions to the time-independent ones. Hence, we are left with a
modified equation in the rotating coordinate system (\ref{z123}).
We have used the Multiple Scale Analysis to find the number of
generated particles. As an example, we have calculated it for the
specific coupled modes $\ds{(1,1,1)}$ and $\ds{(1,2,1)}$,
$\ds{(2,1,1)}$. We have found that the average number of generated
particles increases exponentially in time (\ref{eyusty}). The same
result was obtained for the translational dynamical Casimir effect
problem. In  other words, the formula for the number of generated
particles are the same for both the rotating cavity and the cavity
whose one
mirror moves according to the equation (\ref{L1L2gdh}).\\
Here, we have considered the scalar particle
generation. The vector particle generation can also be studied for
the rotating cavity. This problem is more difficult since the
vector components are mixed as the rectangular cavity rotates. The
generated photons have not fixed polarization as they have in the
translational dynamical Casimir effect problem. So, multiple scale
analysis may not work for that problem since it enable us to find
the number of generated photons in a given polarization.


\begin{thebibliography}{99}
\bibitem{z1} H. B. G. Casimir, Proc. K. Ned. Akad. Wet. {\bf  51}, 793 (1948).
\bibitem{deneyCas} S. K. Lamoreaux, Phys. Rev. Lett. {\bf 78}, 5 (1997).
\bibitem{z2} E. Sassaroli, Y. N. Srivastava and A. Widom, Phys. Rev. A {\bf 50}, 1027 (1994).
\bibitem{z3} C. K. Law, Phys. Rev. Lett. {\bf 73}, 1931 (1994).
\bibitem{z4} O. M´eplan and C. Gignoux, Phys. Rev. Lett. {\bf 76}, 408 (1996).
\bibitem{z9} C. K. Cole and W. C. Schieve, Phys. Rev. A {\bf 52}, 4405 (1995).
\bibitem{z11} V. V. Dodonov, A. B. Klimov, and V. I. Man'ko, Phys. Lett. A {\bf149}, 225 (1990).
\bibitem{z31} V. V. Dodonov, Adv. Chem. Phys. {\bf 119}, 309 (2001).
\bibitem{ekle1} P. Wegrzyn,  J. Phys. B: At. Mol. Opt. Phys. {\bf40}, 2621 (2007).
\bibitem{ekle2} D. T. Alves and E. R. Granhen, Phys. Rev. A {\bf 77}, 015808 (2008).
\bibitem{ekle3} Y. N. Srivastava, A. Widom, S. Sivasubramanian and M. P. Ganesh, Phys. Rev. A {\bf 74}, 032101 (2006).
\bibitem{qw1} Woo-Joong Kim, J. H. Brownell, R. Onofrio, Phys. Rev. Lett. {\bf 96}, 200402 (2006)
\bibitem{qw2} C. Braggio, {\it et al.}, Europhys. Lett. {\bf70}, 754 (2005).
\bibitem{deney1} S. D. Liberato, C. Ciuti, F. I. Carusotto Phys. Rev. Lett. {\bf 98}, 103602 (2007).
\bibitem{z16} A. Lambrecht, M. -T. Jaekel, and S. Reynaud, Eur. Phys. J. D {\bf 3} 95 (1998).
\bibitem{z17} Jeong-Young. Ji, Hyun-Hee. Jung and Kwang-Sup. Soh, Phys. Rev. A {\bf 57}, 4952 (1998).
\bibitem{twomoving1} D. A. R. Dalvit, F. D. Mazzitelli, Phys. Rev. A {\bf59}, 3049 (1999).
\bibitem{twomoving2} M. Ruser Phys. Rev. A {\bf 73}, 043811 (2006).
\bibitem{twomoving3} V.V. Dodonov, J. Phys. A: Math Gen. {\bf 31}, 9835 (1998).
\bibitem{twomoving4} L. Ling, L. Bo-Zang, Chin. Phys. Lett. {\bf 19}, 1061 (2002).
\bibitem{twomoving5} D. F. Mundarain, P. A. M. Neto, Phys. Rev. A {\bf 57}, 1379 (1998).
\bibitem{z5} A. Lambrecht, M. T. Jaekel, and S. Reynaud, Phys. Rev. Lett. {\bf 77}, 615 (1996).
\bibitem{yuceozcak} C. Yuce, Z. Ozcakmakli, J. Phys. A: Math Theor. {\bf 41}, 265401 (2008).
\bibitem{z20} M. Razavy and J. Terning, Phys. Rev. D {\bf 31}, 307 (1985).
\bibitem{z21} G. Calucci, J. Phys. A {\bf 25}, 3873 (1992).
\bibitem{z22} C. K. Law, Phys. Rev. A {\bf 49}, 433 (1994).
\bibitem{z24} V. V.Dodonov and A. B. Klimov, Phys. Rev. A {\bf 53}, 2664 (1996).
\bibitem{z25} A. B. Klimov and V. Altuzar, Phys. Lett. A {\bf 226}, 41 (1997).
\bibitem{z26} Jeong-Young Ji, Hyun-Hee. Jung, Jong-Woong. Park, and Kwang-Sup. Soh, Phys. Rev. A {\bf 56}, 4440 (1997).
\bibitem{z27} M. Crocce, D. A. R. Dalvit, F .D . Mazzitelli, Phys. Rev. A {\bf 66}, 033811 (2002).
\bibitem{z30} M. Crocce, D. A. R. Dalvit and F. D. Mazzitelli, Phys. Rev. A {\bf 64}, 013808 (2001).
\bibitem{z28} R. Schutzhold, G. Plunien, and G. Soff, Phys. Rev. A {\bf 65}, 043820 (2002); G. Schaller, R. Schutzhold, G. Plunien, and G. Soff, Phys. Rev. A {\bf 66}, 023812 (2002).
\bibitem{z29} M. Ruser, J. Opt. B {\bf7}, 100 (2005).
\end{thebibliography}
\end{document}